# X-ray Powder Diffraction from Sub-micron Crystals of Photosystem-1 Membrane Protein

D.A. Shapiro[1,4], H.N. Chapman[5,*], D. DePonte[2], R.B. Doak[2], P. Fromme[3], G. Hembree[2], M. Hunter[3], S. Marchesini[1,4], K. Schmidt[2], J. Spence[2], D. Starodub[2], U. Weierstall[2]

[1]Advanced Light Source, Lawrence Berkeley National Laboratory, Berkeley, CA 94720, USA
[2]Dept. of Physics, Arizona State University, Tempe, AZ 85287, USA
[3]Dept. of Chemistry and Biochemistry, Arizona State University, Tempe, AZ 85287, USA
[4]Center For Biophotonics Science and Technology, Univ. of California at Davis, 2700 Stockton Blvd., Suite 1400, Sacramento, CA, 95817, USA
[5]Lawrence Livermore National Laboratory, 7000 East Ave., Livermore CA, 94550, USA

**Abstract**
We demonstrate that powder diffraction data can be collected from sub-micron crystals of a membrane protein with nearly two orders of magnitude more atoms than the molecules commonly used for powder diffraction. The crystals of photosystem-1 protein were size-selected using a 500 nm pore-size filter and delivered to a soft x-ray beam with a photon energy of 1.5 keV using a dynamically focused micro-jet developed for the serial crystallography experiment at beamline 9.0.1. The 10-micron jet places many such randomly oriented crystals in the x-ray beam simultaneously resulting in a powder diffraction pattern which extends to 28 Å resolution with just 200 seconds of x-ray exposure. The use of the jet for particle delivery allows for a thin sample, appropriate for the soft x-rays used, and continuously refreshes the crystals so that radiation damage is not possible. The small size of the crystals requires the use of lower energy photons for increased scattering strength and increased spacing between powder rings. The powder patterns obtained in this way, from abundant nano-crystals, could be used to provide low resolution molecular envelopes if phased using techniques such as compressive sensing which do not require atomic resolution data. The results also serve to test our aerojet injector system, with future application to femtosecond diffraction in Free Electron X-ray Laser schemes, and for Serial Crystallography using a single-file beam of aligned hydrated molecules.

## 1. Introduction

The major bottleneck in protein structure determination by x-ray diffraction from single crystals is the production of high quality crystals that are large enough for diffraction analysis. Protein crystallography beamlines that utilize third generation synchrotron sources typically require crystals that are at least 20 microns in size for the collection of atomic resolution data. Even micro-diffraction probes are limited to crystals several tens of microns in size, since the high X-ray exposure of small single crystals leads to the severe problem of radiation damage. Most membrane protein crystals suffer X-ray damage during data collection even at cryogenic temperatures. Small single crystals of membrane proteins fail to show high resolution X-ray diffraction after seconds of exposure so data have to be collected using hundreds of partial data sets from many crystals. That method has been used to determine the first high resolution structure of G-protein coupled receptor (Rasmussen, 2007), a class of membrane proteins that represent more than 50% of all current drug targets. Alternatively, data collection has to be performed using large single crystals that can be shifted several times to limit the X-ray exposure at each region of the crystal.

---

* Current address: Centre for Free-Electron Laser Science, Universität Hamburg and DESY, Notkestrasse 85, 22607 Hamburg, Germany

The size of the crystals required for data collection increases with the size of the unit cell. As an example, crystals larger than 500-microns were essential for the structure determination of Photosystem I at 2.5 Å resolution (Jordan, 2001). Photosystem I is a membrane protein with a molecular weight of 1.056 MDa and represents one example of a class of proteins that are difficult to crystallize. The production of large, well ordered, single-crystals of proteins that are difficult to crystallize can take years and may involve very time consuming investigations such as the determination of phase diagrams or seeding techniques. At the same time, optical microscopy evidence suggests that many crystallization solutions may contain large numbers of small crystals of the size of about one micron or smaller (Von Driel, 2007), and the mother liquour in crystal growth wells contains a high concentration of these nuclei. Experiment and theory indicate that such small crystals do not contain enough molecules for statistically significant scattering before destruction of high-resolution detail by radiation damage. These considerations have lead to a revival of interest in powder protein crystallography (Margiolaki, 2008).

In this paper we report preliminary results from a serial protein crystallography project that aims to provide a proof of concept that X-ray diffraction can be detected from a stream of droplets that contains nano-crystals. As the model protein we did not choose lysozyme, but Photosystem I, which is the largest and most complex membrane protein that has been crystallized to date. The central component of this method is a single-file droplet beam, containing nano-crystals, which originates from a Gas Dynamic Virtual Nozzle (aerojet) (DePonte, 2008, Weierstall, 2008) droplet source and traverses a quasi-continuous x-ray beam. The jet may be filled directly with mother liquor from crystal growth cells, allowing us both to test the jet and evaluate the prospects for using it for powder protein diffraction. The current experiment is limited in resolution only by the long wavelength of the x-rays used but the method shows potential for high throughput screening of crystallization drops for micro-crystals and the avoidance of radiation damage during collection of powder diffraction data.

Since the droplets are moving with a velocity of the order of 10 m/s and the x-ray beam is only a few tens of microns wide each protein nanocrystal is exposed to x-rays for only 2 microseconds and thus is not damaged by the ionizing radiation. Using tabulated data for elemental photoabsorption cross sections (Henke, 1993), we obtain a mass absorption coefficient $\mu = 1.2 \times 10^4$ cm$^2$ g$^{-1}$ at an X-ray energy $E = 1500$ eV for a generic protein stoichiometry $H_{50}C_{30}N_9O_{10}S_1$. The energy per unit mass (dose) deposited into a protein nano-crystal during its passage through the X-ray beam can be estimated as $D = \mu E I_0 t = 2.6 \times 10^2$ Gy, where $I_0$ is the incident X-ray flux (number of photons incident per unit area per unit time). This dose is more than four orders of magnitude lower than the Henderson limit, therefore radiation damage is not a concern in our experiments. Large total exposure times are therefore available without damaging the sample because the detector integrates the diffracted intensity as the crystallites are continuously refreshed by the aerojet source.

This new method has high potential for the determination of structures using a stream of micro-crystals synchronized with a pulsed X-ray laser or, in the long term, a stream of oriented and hydrated single protein molecules could be used, which would allow for the study of the diffraction patterns in the transition to the "gas-phase". In this regime, particle-size line broadening prevents identification of the reciprocal lattice and molecular alignment is required. Pending final development of alignment methods for serial crystallography, we show here that X-ray diffraction data may be obtained using our aerojet of small protein crystals without alignment, by applying the methods of powder diffraction to sub-micron crystallites within the droplets. Since the dimensions and symmetry of the unit cell can be determined from the ring pattern, powder diffraction avoids the need to solve the molecular alignment problem. Ultimately this is traceable to the fact that there is a finite number of Bravais lattices because there is a limited number of ways to arrange points periodically in space with identical environments. There is no such constraint for a single molecule.

These diffraction experiments were carried out using 1500 eV x-rays with crystals of the membrane protein Photosystem-1 with sizes of less than 500 nm. The use of such long wavelengths increases the scattering cross-section (thereby decreasing exposure times and particle concentrations needed) and reduces the effects of solution scattering. The protein crystals were size restricted with a nano-porous filter to sizes of around 500 nm. Our preliminary experimental geometry and the long wavelength of the soft X-rays only allows for the observation of the lowest order diffraction rings but our results closely match theoretical predictions from the known structure.

## 2. Experimental

Diffraction data were collected at beamline 9.0.1 of the Advanced Light Source where the beam defining optics are optimized for coherent soft x-ray scattering experiments. The source, a 10 cm period undulator with 43 periods, provides 1500 eV photons in the ninth harmonic, as a Bragg reflection from a plane multilayer mirror, with an approximate photon flux of $10^{11}$ photons per second in a 50-micron focus provided by a monochromatizing zone plate segment. The combination of the multilayer mirror and beryllium vacuum window act as a bandpass filter that removes higher and lower undulator harmonics while the remaining undulator spectrum is dispersed vertically by the off-axis zone plate segment. This 550-micron diameter zone plate segment focuses the third undulator harmonic to a 50-micron spot one inch upstream of the droplet beam. This spot is then selected with a 50-micron pinhole. Experiments requiring greater transverse and longitudinal coherence use smaller beam defining pinholes so that fewer coherent modes may be selected. Accepting the full undulator harmonic results in a bandwidth equal to the source bandwidth that is approximately equal to $1/mN$, where $m$ is the harmonic number and $N$ is the number of undulator periods. This neglects electron beam size and divergence and for our case provides an upper limit on the bandwidth of 0.8%, though the actual value will not be significantly different.

Focusing of the x-ray beam is essential because of the flux-limited nature of this experiment. The beam divergence, dominated by the zone plate focusing with focal length of 80 cm and the large x-ray bandwidth, result in significant angular broadening of the scattered peaks. An undulator source at the ALS has a central cone divergence of approximately 40 μrad, which is at least a factor of 20 smaller than the other sources of beam spread. At the largest scattering angles recorded (17 degrees), the angular width of a scattered beam is at least 0.13 degrees though the precise value is not known without a quantitative measurement of the x-ray monochromaticity. The detector used, a Princeton Instruments MTE2-1300B CCD, has sufficient angular resolution to make accurate measures of the powder peak profiles. The 20-micron pixels, at a 5 cm working distance, provide an angular resolution of 0.02 degrees or 6 detector pixels per peak FWHM. Therefore, the detector was binned by a factor of two in each dimension for faster readout.

The particles of interest are delivered to the x-ray beam in a continuous liquid jet which breaks into droplets further downstream due to a Rayleigh necking instability (Rayleigh, 1878). The jet, which is described in detail elsewhere (Weierstall, 2008), consists of a column of particle-containing solution which flows through a 50 micron inner-diameter hollow tube (fiber optic) and is accelerated upon exiting the tube by a coaxial flow of carbon dioxide gas into vacuum. The $CO_2$, which acts to focus the liquid stream, condenses as dry ice on a liquid-nitrogen cooled coldtrap, thereby minimizing the rise in pressure caused by the liquid and gas leak into the vacuum chamber. The acceleration by co-flowing gas thus serves the purpose of reducing the jet diameter while allowing use of a larger nozzle, which is less likely to clog. A typical jet diameter for the images shown here was 10 microns, which corresponds to a flow rate of a few tens of micro-liters per minute. Figure 1(A) shows a liquid jet emerging from the nozzle exit, averaged over an exposure time of 1 second, while figures 1(B) and 1(C) show flash (nanosecond) images of the droplets some greater distance from the nozzle tip. In figure 1(B), we have controlled the droplet generation rate using a piezoelectric actuator, which may be synchronized to a pulsed X-ray source for experiments aimed at reading out the diffraction pattern from single molecules, using a very fast read-out area detector (Neutze, 2000, Ourmazd, 2008). However, for the results reported here, the droplet breakup was un-triggered and occurred with the intrinsic Rayleigh instability frequency.

## 3. Results and Discussion

Photosystem I (PSI, 1JBO in PDB, a=b=288 A, c = 167 A, space group P63, 1 MDa, 72,000 atoms, two trimers per unit cell, 78% solvent) crystal samples were prepared from the cyanobacterium *Thermosynechococcus elongatus* as described elsewhere (Jordan, 2001). Briefly, the cells were grown under low-light conditions to allow for a high yield of trimeric photosystem I. The cells were harvested, lysed, and the proteins extracted with detergent (beta-dodecylmaltoside) and isolated using an anion-exchange chromatograph. The eluent solutions were composed of 20 mM **MES** pH 6.4, 0.02% (m/v) (beta-dodecyl maltoside), and a $MgSO_4$ gradient was used for the elution of the protein. After purification, the photosystem I was diluted to low-salt conditions (final concentration of 20 mM MES pH 6.4, 6 mM $MgSO_4$ and 0.02% (m/v) beta-dodecyl maltoside), which leads to the growth of micron and sub-micron sized crystals spontaneously overnight at 4°C.

The size of the PS1 crystallites was then restricted by placing a 500 nm filter in the liquid line of the aerojet. With many such crystallites per droplet, the recorded diffraction pattern contains contributions from crystallites in all possible orientations. This generates a sphere centered on the origin of reciprocal space for every reciprocal lattice point, which is intersected on a circle by the Ewald sphere, producing the familiar Debye-Scherrer diffraction rings. In reality, this sphere is a shell of finite thickness, which depends on several experimental factors. These include beam divergence, spectral bandwidth, crystallinity effects such as imperfections or finite crystal size, and instrumental broadening resulting from finite angular resolution. Fig. 2a shows the powder diffraction profile up to a scattering angle corresponding to 28Å resolution. As noted earlier, the peak width due to the characteristics of the x-ray illumination is 0.13 degrees. In figure 2, the measured powder peaks have FWHM widths of about 0.25 degrees. The additional peak width must therefore result from crystallinity and finite size effects. If the small crystallites are coherently illuminated then the Bragg peaks will be convolved with the diffuse coherent scatter that represents the overall crystal morphology. Averaging this effect over many such crystals will give a broadened powder profile whose width is inversely proportional to the average crystal size. The number of coherent modes in our beam is given approximately by

$$N = \frac{2x \cdot x'}{\lambda},$$

where $x$ is the beam size in the monochromator focus and $x'$ is the beam divergence. This gives 27 coherent modes in our 50-micron beam or about 0.6 microns per mode. Thus, all the crystallites are coherently illuminated and it is not likely that neighboring crystals interfere coherently. The exposure time for the diffraction pattern was only 200s, with a flow rate of 1μl/s, i.e. only very small amounts of samples (0.2ml) have been used for the diffraction pattern. This small sample volume would allow the development of the use of the droplet stream for the high-throughput robotic screening of crystallization drops for micro-crystals that are not visible by optical microscopy.

The projection of three-dimensional data onto one dimension in powder diffraction results in a loss of information in two ways (see David, 2002 for a review). First, at large angles the decreasing ring spacing eventually becomes less than the angular resolution of the instrument, setting a limit on the ultimate resolution of the density map. Second, overlaps can occur in space-groups of higher symmetry than orthorhombic due to indexing accidents, so that structure factors of different magnitude contribute to the same ring. Degeneracies due to point, space-group and chiral symmetries may also be masked, and these must be known to extract structure factor magnitudes unless, as in the charge-flipping method, the analysis is always performed in P1. In a previous paper (Wu, 2006), we have shown how a variant of this charge-flipping algorithm can resolve both the accidental degeneracies and the symmetry degeneracy in powder data from inorganic crystals. The most common space group for proteins is orthorhombic P 2/1 2/1 2/1, which has no accidental overlaps. The density of reciprocal lattice points for a triclinic crystal may be found as a function of scattering angle $\theta = 2sin^{-1}(\lambda/2d)$ (twice the Bragg angle) as (David et al 2002),

$$\frac{\Delta N}{\Delta \theta} = \frac{16\pi V sin^2(\theta/2)cos(\theta/2)}{\lambda^3}$$

Thus peak density is proportional to the square of the scattering angle for small angles, inversely proportional to $d^2$ (for $d<\lambda$), proportional to cell volume, and inversely proportional to $\lambda^3$. This function has a maximum at $\Theta_B=55°$, where $\Delta N \approx 19V$ per degree for hard X-rays around $d=1Å$. Our use of long wavelength (0.8 nm) thus greatly improves peak separation, while limiting ultimate resolution (an upgrade to 4Å wavelength is planned). This improves angular resolution due to beam divergence, but not due to particle-size broadening, which is independent of beam energy. We use an undulator insertion device for high flux (focused to the 20 micron diameter of our droplet beam), and take advantage of the $\lambda^4$ dependence of coherent scattered intensity (Howells, 2008). This enables short exposure times from our droplets, which contain relatively few protein molecules. The solution of such large proteins by powder methods at low resolution may best be attacked by using a combination of parameterized theoretical modeling, molecular replacement, and/or isomorphous replacement.

## 4. Conclusion

We have demonstrated for the first time solution x-ray scattering from a micro-jet of nano-scale crystallites of a membrane protein using a synchrotron undulator source. For non-crystallized particles, this procedure yields the equivalent of a small-angle x-ray scattering (SAXS) pattern while for small crystallites it yields a powder diffraction profile. Our analysis shows that quantitative structural information may be obtained from the scattering data acquired in this manner. The primary benefit of this technique is that the smallest available protein crystals may be used without any concern for radiation damage in a windowless geometry which provides an optical path short enough for use of medium-energy X-rays. Temperature variation along the jet may also be used to aid in peak deconvolution. Currently, there is no other x-ray scattering technique that can utilize such small crystals with such ease. Furthermore, shorter wavelengths and the possible use of particle alignment schemes, currently under development, may allow for the *ab initio* phasing of these diffraction patterns and provide structural maps in the several angstrom range. The incident flux density required for x-ray imaging is inversely proportion to $\lambda^2$ while the dose (energy deposited per unit mass) is relatively independent of energy from 1-10 keV. This seems to indicate that the lowest energies compatible with the desired resolution should be used. Recent theoretical simulations show that 7 Å resolution should be obtainable using planned undulator sources with energies of around 3 keV.

The aerojet droplet source, being a very high brightness particle beam, is also ideally suited for experiments with pulsed x-ray sources. The triggered droplet beam may be precisely synchronized to the x-ray pulses so that all proteins are utilized. This is in contrast to electro-spray injection schemes that produce a highly divergent particle beam and cannot be synchronized. Furthermore, the remaining water layer surrounding the proteins may be continuously adjusted by altering the transit time, i.e. evaporation time, between droplet source and x-ray beam. It has been suggested that this water layer may act as a tamper to slow the explosion of proteins in such intense x-ray pulses and allow the use of longer pulses for the acquisition of atomic resolution diffraction data (Hau-Riege, 2007). Finally, this aerojet device produces a stream of uncharged particles and thereby eliminates possible damage to the molecules associated with the high molecular charge as observed in electrospray mass spectroscopy.

**Acknowledgements**

We graciously acknowledge support from the staff of the Advanced Light Source at Lawrence Berkeley National Laboratory. This research is supported by grants from The Center for Biophotonics Science and Technology at the University of California at Davis, the National Science Foundation (IDBR-0555845) and ARO (W911NF-05-1-0152). The Advanced Light Source at Lawrence Berkeley National Laboratory is supported by the Director, Office of Science, Office of Basic Energy Sciences, Materials Sciences Division, of the U.S. Department of Energy. The structural project on Photosystem I is supported by the National Science Foundation, grant number 0417142.

**Figures**

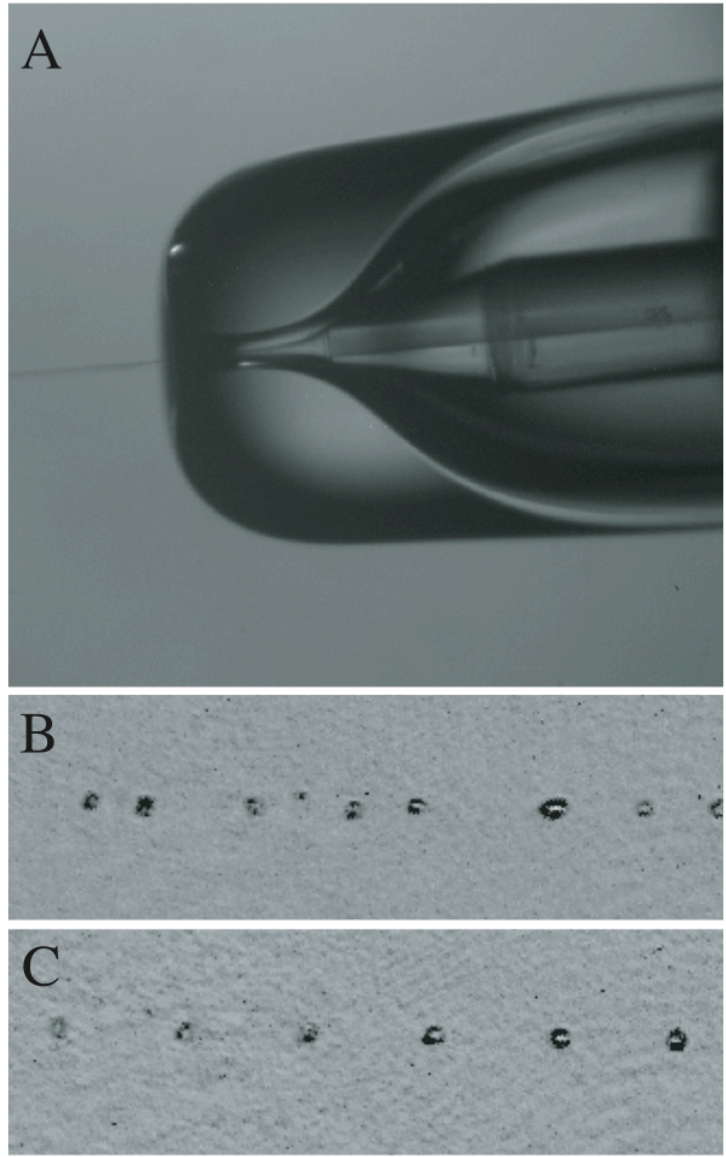

**Figure 1** (A) Aerojet droplet source. An inner capillary, a hollow fiber-optic with 50 micron inner diameter, carries the particle solution which is pressurized to 150 PSI. The outer glass capillary carries the focusing $CO_2$ gas at 15 PSI, which expands into vacuum producing a small jet of solution that breaks up into droplets. (B) Flash image (100 ns exposure time) of the untriggered droplets which are unevenly space and have a distribution of sizes. (C) Triggered droplets are evenly spaced and have a much narrower size distribution.

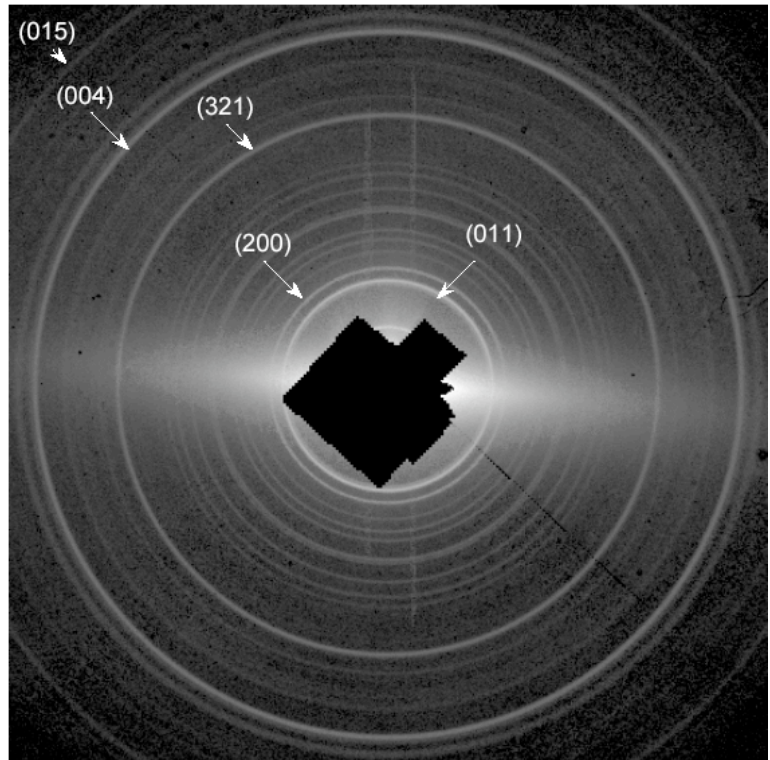
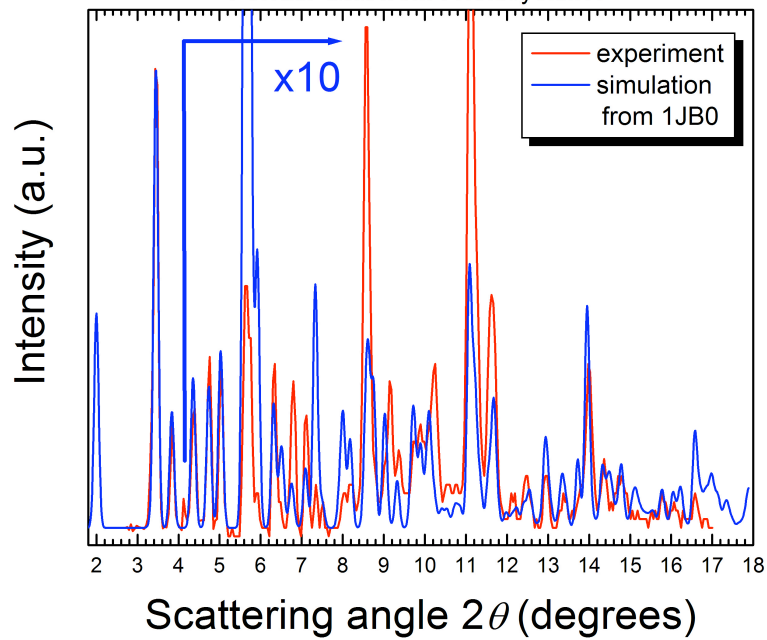

**Figure 2** Powder diffraction pattern from 500 nm crystals of photosystem 1 membrane protein. The total exposure time is 200 s with about $10^{10}$ photons/s/micron$^2$ incident on the 10 micron diameter jet. The peak positions match well with the peaks calculated from the known structure (PDB 1JB0) though differences in intensity can be seen.

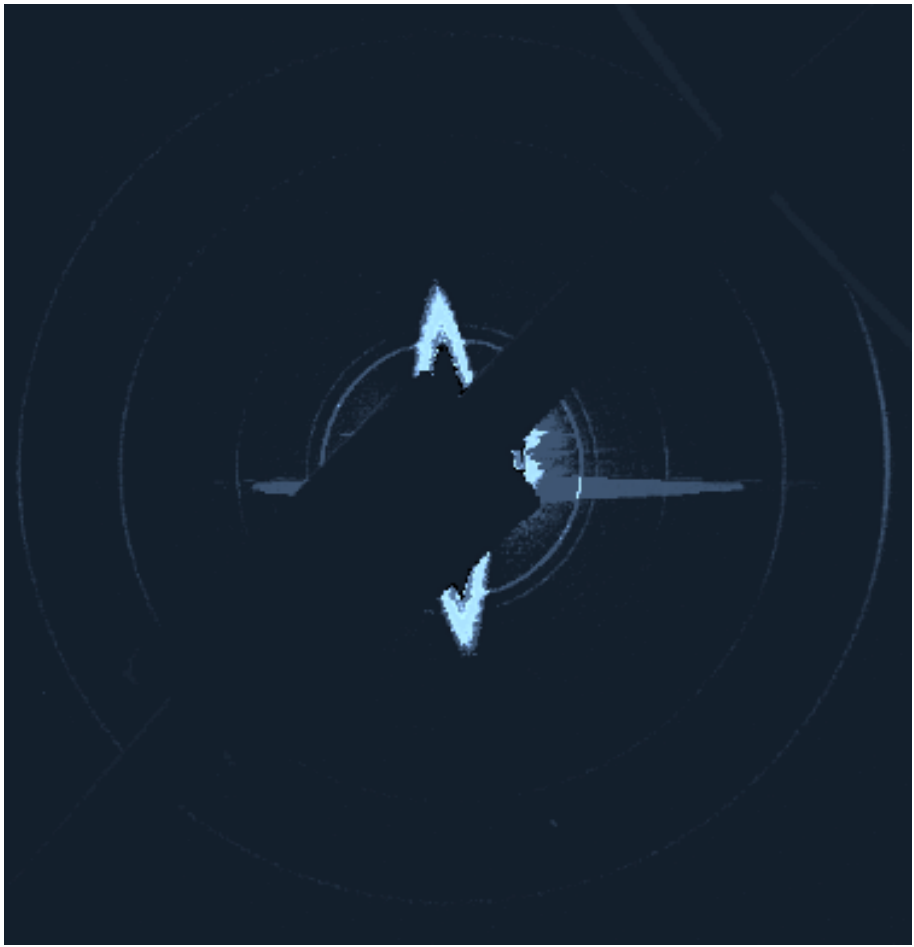

**Figure 3** Running the jet with a 10 micron filter, i.e. allowing larger crystals through, shows evidence of flow-alignment. The above pattern is the difference pattern between data collected using the 500 nm filter and that collected using a 10 micron filter. With the jet flowing left to right, arcs of intensity can be seen that indicate long crystals have aligned with the liquid flow. Surprisingly most of the intensity in the pattern is due to the 500 nm crystals.